# An economical in-class sticker microfluidic activity develops student expertise in microscale physics and device manufacturing

Priscilla Delgado[1,2,3], C. Alessandra Luna[1,2,3], Anjana Dissanayaka[1,2,3], Oluwamayokun Oshinowo[1,2,3], Jesse J Waggoner [4,5], Sara Schley [1], Todd Fernandez [1], David R. Myers[1,2,3,*]

[1] The Wallace H. Coulter Department of Biomedical Engineering, Georgia Institute of Technology & Emory University, Atlanta, GA, USA

[2] Department of Pediatrics, Division of Pediatric Hematology/Oncology, Aflac Cancer Center and Blood Disorders Service of Children's Healthcare of Atlanta, Emory University School of Medicine, Atlanta, GA, USA

[3] Parker H. Petit Institute of Bioengineering and Bioscience, Georgia Institute of Technology, Atlanta, GA, USA

[4] Emory University, Department of Medicine, Division of Infectious Diseases, Atlanta, Georgia, USA

[5]Rollins School of Public Health, Department of Global Health, Atlanta, Georgia, USA

* Corresponding Author: david.myers@emory.edu (ORCID: **0000-0001-9659-3274**)

**Abstract:** Learning miniaturization science remains challenging due to the non-intuitive behavior of microscale objects and complex assembly approaches. Traditional approaches for creating microsystems require expensive equipment, facilities, and trained staff. To improve as well as democratize microdevice education, we created a new educational activity that enables students to build and test advanced microfluidics by leveraging "sticker" microfluidics, composed of double-sided dry film adhesive layers. Along with T-mixers and bubble generators, our activity is the first to enable students to build a valve and F-mixer in the classroom setting. This helps emphasize less intuitive aspects of device manufacturing such as the creation of complex 3-dimenstional shapes with layers and layer alignment. In addition, this paper provides the first reported quantitative data on significant improvements in student knowledge and confidence from building and testing several common devices. All 11 students had substantial improvements in conceptual mastery and confidence after the activity. Student responses to a guided reflection highlight how the activity supports a variety of learning needs and preferences. Given the impact on student learning, our low-cost activity helps reduce global barriers to miniaturization science education.

**INTRODUCTION**

Microsystems have significantly impacted the world around us, and are found in smartphones [1], vehicles [2], clinical diagnostics [3], and cutting edge research [4]–[9]. Microfluidics in particular have significantly impacted point of care diagnostics given their low cost, portability, automation of complex lab processes, and reduced equipment needs [3], [10], [11]. As objects shrink to the scale of microns, concepts of motion and behavior fundamentally deviate from our experiences working with objects on the scale of centimeters to meters. A well-known example of difficulty in transitioning macro to microfluidic is microscale mixing. Student experience difficulty because macrofluidic mixing is dominated by turbulent flows, a phenomenon that is absent in the diffusion-dominated mixing of strictly-laminar microfluidics [12], [13]. Similarly, the layer-by-layer processes used to manufacture microscale devices bears little resemblance to techniques common in macro-fluidics, such as pipe assembly, machining, or extrusion/forging.

Taken together, the non-intuitive physics and manufacturing of microdevices have been shown to make it difficult for students to relate microdevice concepts to prior knowledge and experiences. Referred to within learning sciences as 'misconceptions' these failures of far transfer of prior knowledge make learning and assessing knowledge difficult [12], [13]. With these types of challenges, creating effective learning experiences [14] related to microsystems that rely on a traditional lecture-based approach remains challenging. Previously identified strategies use a "deep approach to learning" and focus on experiences that force students to confront their misconceptions [13], [15].

Laboratory and hands-on activities offer an effective method to supplement traditional lecture to create such learning experiences. Unfortunately, microsystems-focused activities can be difficult to implement in the classroom setting due to challenges in getting access to microchips and microdevices suitable for teaching [12]. Further, given the cost of tools, facilities, and expertise needed to create microdevices, many educational institutions do not have the resources for hands-on microsystems fabrication and student laboratories. The result is that access to the tools for effective microfluidics education is severely limited, especially in low resource settings where microsystems advantages could prove most transformative.

Some examples of techniques used to increase access to classroom based microfluidics and microdevices have used a variety of materials including paper [16], Jell-O [17], chocolate [18], shrinky dinks [19], and acrylic paired with double sided tape [20]. These activities have key advantages in that they are low cost, use readily available materials, and provide a visual demonstration of key microdevice concepts that students struggle with. Learning activities have been implemented in numerous formats and age groups including middle [20] and high school students [19], [21], undergraduate problem-based courses [22], to mini-course modules for MS/PhD students that mix multiple pedagogies [23]. Feedback from students and teachers on these activities has generally been positive [18], [20], [22], [24]. These existing activities lay the groundwork for more advanced manufacturing work by having students create single layer devices or multilayer serpentine channels [20].

Our work complements and extends the excellent body of work in microfluidics education with several key innovations. First, our activity includes well known microfluidics (Fig 1a-b) but now adds advanced functional multi-layer microdevices, specifically geometrically complex mixers (Fig 1c) and valves (Fig 1d) to emphasize layer-by-layer assembly of complex 3-dimensional devices, tolerances, alignment, mechanical motion (valves), and splitting flows to enhance diffusive mixing. Such devices have not been created in the classroom previously. We accomplished this by leveraging previously reported methods to construct individual layers of multilayer microfluidics using laser cut dry film adhesives (double sided tape) [25], [26]. Our second innovation is reporting most comprehensive quantitative data to date on student microdevice learning, student confidence, and the impressions that students recalled 1 week later. We specifically were able to measure student knowledge before and after completing the activity. Importantly, our activity uses economical materials (Supplementary Table 1&2), is highly portable, is safe (Fig 1e), and can be easily visualized by eye (Fig 1f,g) (Supplementary information, Supplemental Videos 1 – 4) in the classroom setting.

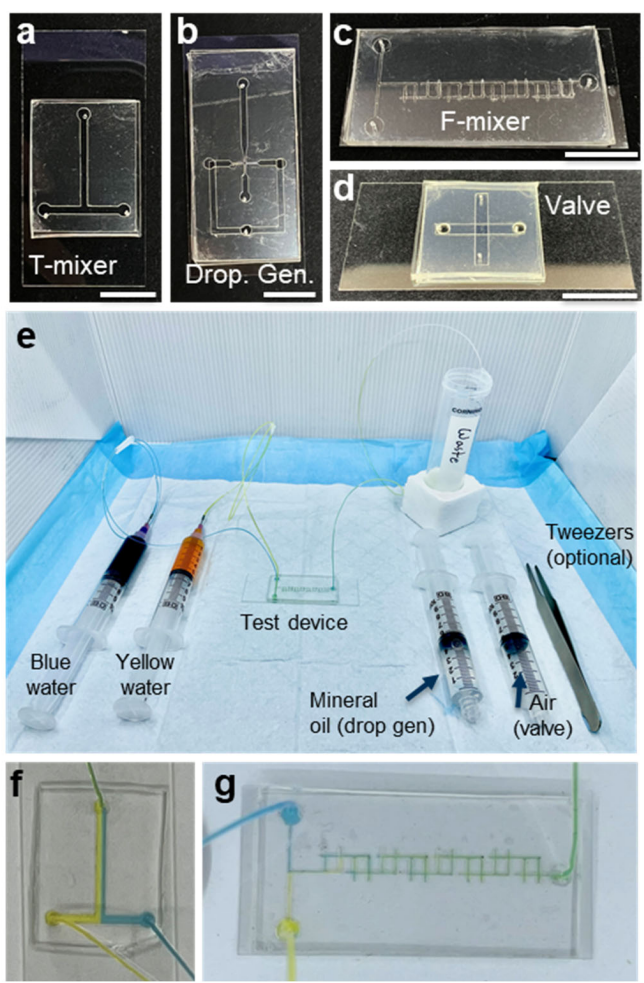

**Fig 1 Economical sicker microfluidics can be assembled and tested in-class to create a significant learning experience related to microscale physics and microdevice manufacturing.** a-d) Our activity features 4 devices of increasing complexity to enable students to build confidence e) Students can operate the devices with gravity fed or hand actuated syringes with colored water and mineral oil. All fluidic testing components are inexpensive and can be reused to reduce costs (Supplementary Table 2). c) All devices operate as expected and

can be visualized with the naked eye, and recorded with a smartphone camera (ESI includes video demonstrations) (scale bar is 1 cm)

**METHODS & MATERIALS**

**Demographics & Context**

This activity was performed during the 4[th] week of the Georgia Institute of Technology Wallace H. Coulter Department Biomedical Engineering Course "Translational Microsystems". During the first three weeks of class, the following topics were covered in a more traditional lecture format: biomedical microsystems, scaling laws, photolithography, soft lithography, surface micromachining, microcontact printing of proteins, and advanced microfluidics. Each lecture used a handout in which both student and instructor worked together to fill in key information. The activity we focus on in this paper then occurred during the fourth week. Within the class, all 11 students permitted their data to be used in the study, 2 undergraduates and 9 graduate students. Of these students, 3 had no prior experience with microdevices, 3 had 0-1 years of research experience with microdevices, and 5 had more than 1 year of research experience with microdevices.

**Activity description**

The activity we created occurred over two consecutive days in the class. On the first day, teams of students assembled several microfluidics devices. On the second day, the teams then tested the devices they had created. Each course day lasted 75 minutes. We designed the activity to promote both affective and cognitive practices to promote an effective learning experience [14]. Affective practices include those non-cognitive aspects that impact learning, while cognitive practices promote the self-analysis of learning and connections between subject areas. To that end, our activity is intended to build motivation and engagement by enabling students to interact with devices in a hands on, self-constructed way. This affective practice arouses interests and provides a stimulating complement to classroom lectures. From a cognitive standpoint, feedback on the device operation is timely, and the student teams promote meaningful peer-interactions. In addition, when offered as a part of lecture-based course, as is the case here, students must integrate their knowledge and skills from lecture to create an actual working device.

*Device design and construction*

Each device is composed of substrate, such as glass or acrylic, several laser cut layers of dry film adhesive (i.e. double sided tape), and a fluidic interface layer composed of polydimethylsiloxane (PDMS) (Fig 2a). The devices and device layers were sized such that they all experienced phenomena associated with microscale structures, but remained large enough to be easily visualized with the naked eye. To assemble the devices, students peel off the backing layers from the double-sided tape layers and stack them together. While each layer must be carefully aligned to the previous layer to ensure proper device operation, the transparent nature of each layer greatly simplifies this process and subsequent troubleshooting. To aid in alignment, the tape layers and PDMS fluidic interface layer had the same rectangular outline, such that students could also align the outside edges of each layer if preferred.

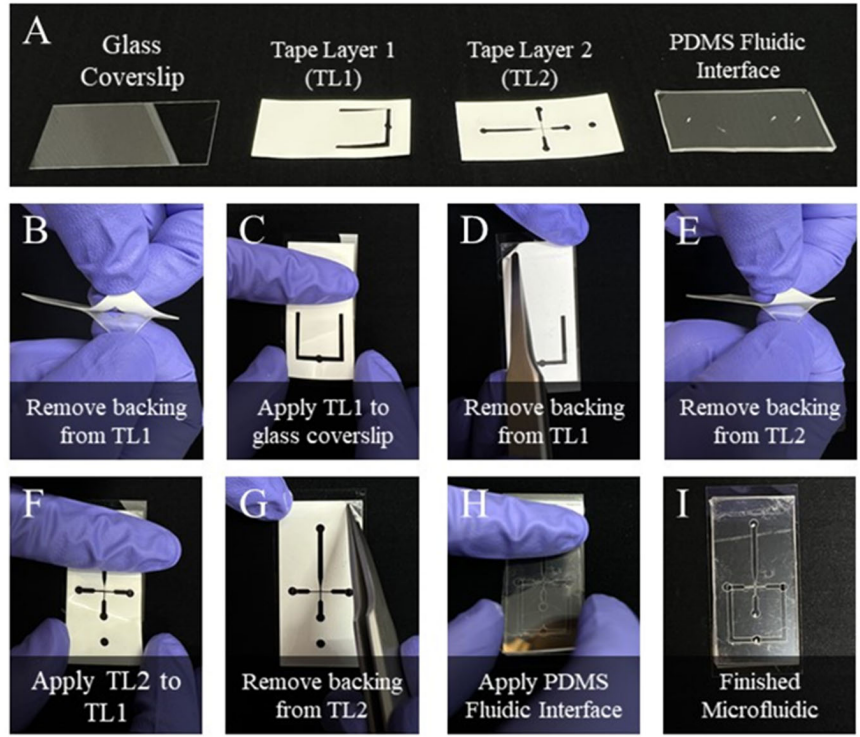

**Fig 2 Each team received step-by-step visual instructions for each device as shown by this example.** a) All step by step instructions begin by showing all the key layers of the device followed by b-h) visual instructions of the remainder of assembly and finishing with i) a final top down view of the device. To reinforce and enhance the learning associated with assembling the microfluidic, handouts were provided for each device asking students to draw the 3d structure test the operation.

A key strength of our approach is the use of fluid connections that are interference based. Since the hole in the PDMS layer has been slightly undersized relative to the tubing, and since PDMS is a flexible material, students simply push the tubing into the holes to create a sealed connection. This enables students to easily change fluidic connections, which proved to be especially useful and many students needed multiple attempts to find the correct fluidic configuration for the more advanced devices, such as the droplet generator.

The high adhesive strength of the tape helps facilitates device construction and can provide an important learning opportunity. While it facilitates bonding and prevents leaking between device layers, each layer can only be applied once and cannot be removed if misaligned. This mirrors what occurs with traditional microdevice fabrication, where misalignment of layers cannot be easily repaired and the misaligned devices must be discarded. Hence, via experience, students learn why precision is needed when adding additional microdevice layers since devices with misaligned layers can not be salvaged. In anticipation of mis-assembled devices, extra kits for each device were available for each team. We also created fully assembled devices, which could be reviewed by each team at any time, and later used if assembly of their device was not successful.

*Instruction provided to students for device assembly*

On the first day of the activity, students were placed into teams composed of two or three individuals, and provided with kits that included all necessary materials for device construction as well as handouts. The handout and kits were designed to encourage teams to construct devices in an independent manner at their own pace. Instructors provide support, help troubleshoot, but do not lead the activity. As such, students have the opportunity to learn from errors, such as misassembled devices, misaligned layers, or inadvertently wrinkled layers that do not provide a water tight seal.

The accompanying handouts (ESI) include step-by-step instructions and pictures on how to assemble each device, with an example shown in Fig 2. The handout encourages teams to construct the devices in the following order of complexity, beginning with the T-mixer (3 layers), then the droplet generator (4 layers) and F-mixer (4 layers), and finally the valve (5 layers). The handout also includes questions related to device assembly and manufacturing to help reinforce skills related to manufacturing, alignment, and translating 2-dimensional drawings of layers into 3-dimensional devices. To assist with assembly, some materials such as scotch tape and tweezers were also available to the whole class.

*Instruction provided to students for device testing*

On the second day of the activity, the same student teams tested each microfluidic device using syringes, ancillary tubing, mineral oil, and water with food safe dyes added. The same accompanying handout also includes instruction related to testing the devices with accompanying questions. However, the handout provides less explicit instructions than with assembly. Instead, the needed syringes are simply listed with a written description of the desired outcome. Students could choose to use either gravity fed fluid flow or hand actuated syringes to feed the mixing devices. To induce gravity fed flow, students could either hold an open syringe partially filled with water above the device, or place the device in a ring stand. This more open yet structured format invites experimentation, observation, and reflection that helps students learn from misconnected fluidic ports. Instructors provided support with troubleshooting and also invited teams to collaborate with each other in cases where one team was having difficulty with a particular device.

To facilitate comparison between mixers, the handout encouraged the students to test their devices in a different order than assembled (T-mixer, F-mixer, droplet generator, valve). In the T-mixer, students added yellow and blue water and can observe that the fluids do not mix (Supplemental Video 1). In the F-mixer, students will see the water gradually turn green as the blue and yellow inlet fluids mix (Supplemental Video 2). We note for instructors that hand actuation syringes worked best for the droplet generator, specifically to suspend blue water droplets in mineral oil (Supplemental Video 3). Similarly, hand actuation worked well for students actuating the valve, and enabled them to understand how pressure differentials between the fluidic channel and valve channel modulated flow (Supplemental Video 4).

*Learning Objectives:*

As a result of completing this activity, students should be able to do the following:

- Understand the difference between micro and macroscale mixing, specifically laminar versus turbulent mixing
- Understand how split and recombine mixers operate
- Understand how multiple microdevice layers can assemble into a complex 3d shape
- Successfully construct a multilayered microfluidic device whose operation depends on correct alignment

- Understand how microfluidic droplet generators work, specifically how geometry and the use of immiscible fluids can be harnessed to create uniform droplets
- Understand the operation of microfluidic valves
- Understand how air bubbles influence microfluidic device operation
- Understand how tolerances facilitate the assembly of microfluidics
- Draw the top view and cross-sectional profiles of a T-mixer, F-mixer, droplet generator, and valve
- Correctly connect liquid, air, and oil inputs to achieve a desired outcome
- Perform basic microfluidic troubleshooting with regards to identifying leaks and operation limits

**Activity assessment**

*Data collection*

The effectiveness of the in-class activity was evaluated using a pre-post knowledge test developed for this activity (Supplemental Materials) as well as a guided reflection administered 1 week after the activity. The knowledge test was composed of 9 open ended short answer questions with topics broadly related to microfluidic scaling physics, operation of microfluidic valves, fluid mixing at the microscale, droplet creation, and the effects of air bubbles in microfluidics. The manufacturing questions asked students to draw cross section and top views of a mixer, valve, and droplet generator, in addition to asking students to reflect on the role of tolerances in microdevice manufacturing. As part of the knowledge test, students were asked to rate their confidence in their answers on a scale of 1 (least) to 5 (most confident). Each question was scored by the course instructor on a scale of 0 to 10, with a maximum score of 90.

In addition to the objective knowledge tests, we asked students to record the subjective learning experience in two ways. First, we included a prompt in the post-test asking students to self-report what they learned. After 1 week, students were asked to complete a structured reflection consisting of four open-ended questions: 1) Which devices did you understand better? 2) Which devices are you more confused about? 3) What are three things that you learned that you found most important? 4) How did your understanding of microfluidic mixing change?

*Data Analysis*

For data analysis, we report confidence weighted scores for the knowledge test rather than confidence and accuracy score separately. Confidence weighted scores were calculated by multiplying the score for each question (0-10) by the student's confidence in their answer (1-5). Confidence weighting provides a more holistic assessment of conceptual mastery, defined as both a correct answer and recognition that the answer is correct [27]. Changes in confidence weighted scores were tested at the 0.05 significance level using the Wilcoxon signed-rank test. All data on student scores and confidence were analyzed with OriginPro (OriginLab Corp). The supplement contains figures replicating the analysis with score and confidence separated.

All open-ended self-identified learning responses were evaluated and categorized into the following categories: New design and physical insights, setup and troubleshooting areas, layer by layer design learning, new fabrication methods, significant positive experience, and the societal need for similar devices. These categories emerged from the data and were generated by the course instructor. All student comments are available in the supplement information (Supplementary Table 3 & 4).

**RESULTS AND DISCUSSION**

**Confidence weighted scores significantly increase after the activity**

After completing the activity, the students' confidence-weighted scores related to microdevice physics significantly increased. The pre-assessment knowledge test helped establish both the knowledge level and confidence of the students understanding of concepts covered. There was a broad spread in the student's pre-test confidence weighted scores for both new concepts and for concepts covered in the course (Fig 3). We interpret the broad spread as likely reflecting the noted limitations of lecture-based approaches to microfluidics concepts as well as variation in participants' prior experience with microfluidics. The pre-assessment data also showcased the variability in the student's knowledge across different concepts, with a moderate degree of understanding in some areas (e.g., air bubbles and scaling), but lower expertise in others (e.g., valve physics and mixing) (Fig 3). The overall pre-assessment confidence on each question was 2.6 out of 5, and there was a poor relationship between confidence and knowledge (Supplemental Figure 2). In an effort to provide a rigorous analysis, we also examined student scores and student confidence independently. Student confidence alone significantly increased for every concept, and student knowledge increased for all concepts, although the concept of air bubbles did not see a statistically significant increase.

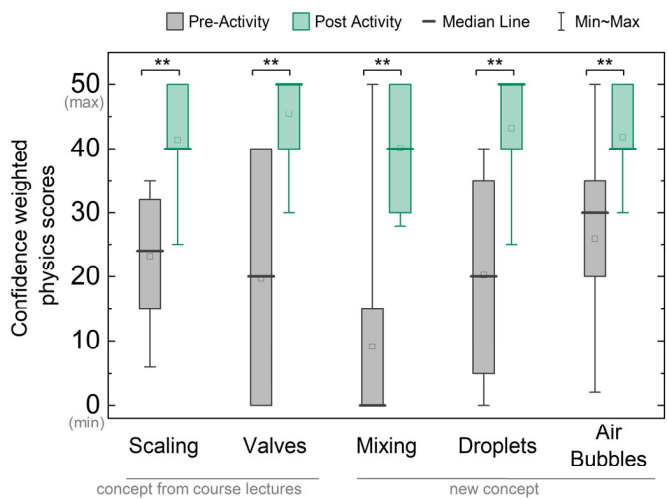

**Fig 3 Students experienced significant increases in confidence weighted physics scores after completing the activity.** Questions for the assessment included both concepts from lecture as well as those outside of lecture learned from completing the activity. The higher assessment scores related to new concepts were also associated with students who had prior microfluidic research experience. (** denotes $p < 0.01$)

The activity promoted a larger increase in student expertise of microdevice manufacturing as compared to physics concepts (Fig 4). A one-way analysis of variance (ANOVA) revealed that there was no statistical difference between physics and manufacturing post-assessment confidence weighted scores. However, student confidence weighted pre-assessment scores related to manufacturing were lower than the pre-assessment scores for the physics questions.

We presume that there could be multiple reasons drive the lower pre-assessment scores related to manufacturing. First, is the availability of resources to develop an understanding of physics and manufacturing. Although macroscale physical phenomena do not match microscale behaviors, students interested in this topical area can get access to media and other educational resources that convey many of these concepts. Unfortunately, fewer resources are available that help convey concepts related to layer-by-layer construction of 3D shapes and spatial reasoning. Moreover, the resources that are available can still be difficult for students to internalize into a generalizable expertise. Another possible explanation is that students may have prior coursework in physics and less in manufacturing.

Our activity promotes expertise in this area by asking students to organize layers and construct multi-layer working devices whose operation depends on successful alignment. Students directly observe how planar shapes assemble into more complex 3D shapes and how tolerances ease assembly. Overall, the resulting high confidence weighted post-assessment scores (Fig 4) suggest the activity played a key role in developing expertise in manufacturing concepts. As before, we also examined student knowledge alone and student confidence independently to increase the rigor of our approach. Student knowledge and confidence in manufacturing concepts all increased in a statistically significant manner. While it is not clear whether the activity alone or whether the activity synergized with traditional lectures, it is clear that offering the activity in the context of a microsystems course can play a key role in the development of expertise.

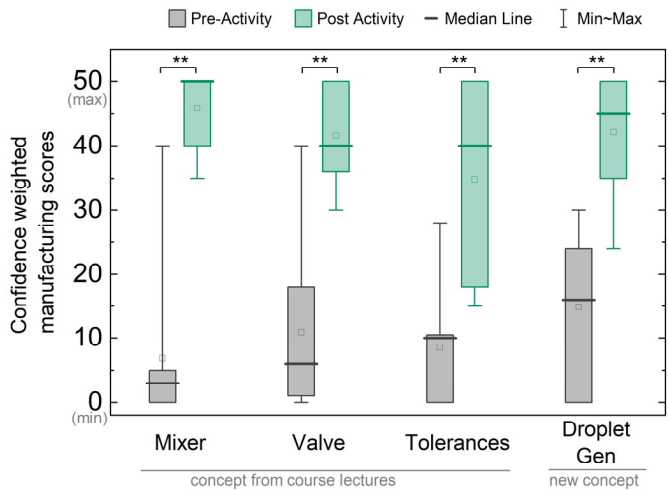

**Fig 4 Students experienced significant increases in confidence weighted manufacturing scores after completing the activity.** Manufacturing pre-assessment scores were lower on average as compared to physics based questions, despite being mostly discussed in class. After completing the activity, post-assessment scores indicated mastery of these concepts. (** denotes $p < 0.01$)

We also examined how individual student performance varied when stratified by degree program and years in a PhD: seeking BS or MS; in PhD Yr 1, in PhD Yr 2+. The students in the BS or MS program did not have prior experience with microdevices, whereas the number of years in a PhD served as a general indicator of the number of years of research experience with microdevice. As expected, PhD students in year 2 or greater had statistically higher weighted pre-assessment scores then students in the BS/MS program, which can be attributed to their prior experience. Regardless of degree program, all students demonstrated significant improvements between pre and post assessment scores after completing the activity (Supplemental Figure 1).

**Student reflections indicate activity supports individualized development of mastery**

Student self-identified improvements in device understanding one week after the activity correlated with device complexity (Fig 5a), highlighting the importance of including more advanced microfluidic devices. The T-mixer has a single fluidic layer in which two fluid streams meet in the middle, and 60% of students commented on understanding this structure better. The F-mixer was most noted by students in 90% of comments.

In an effort to gather feedback on opportunities for improvement, especially related to student learning, our guided reflection included a prompt in which students could comment on whether the activity increased their confusion. All students reported no additional confusion related to the devices built and testing in the activity (Fig 5B). One notable comment mentioned that the learner was more confused about traditional fabrication approaches. Although the exact reason for this confusion is unknown, it is tempting to speculate that performing the activity may have exposed misconceptions related to traditional fabrication.

Student self-identified learning demonstrates that our activity supports the development of expertise in diverse backgrounds. To better understand what concepts were the most important to students, the self-identified learning responses were categorized into groups (Fig 5C). Categories with the highest response rates included concepts related to device manufacturing, physics, design, and troubleshooting. It is worth nothing that many of these comments highlighted non-cognitive learning aspects of microdevices as important, such as understanding the societal need for microdevices and having a significant positive experience with microdevices. The key lessons learned identified directly after the activity (Supplementary Table 3) and 1 week later (Supplementary Table 4) were similar, although later comments tended to be more sophisticated and specific. However, this could be due to the difference in the prompt ("what did you learn?" vs. "what three things did you find most important?"). The similar responses suggest that the activity was impactful as students were still able to identify key lessons learned 1 week after the activity.

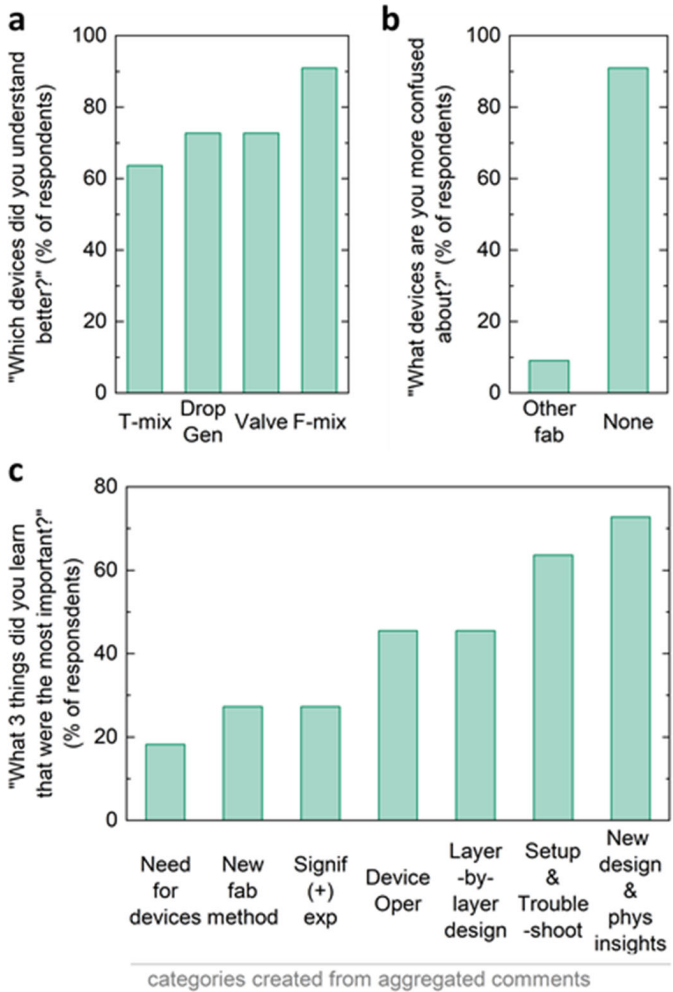

**Fig 5 Responses to open-ended short answer questions asked 1 week after the activity suggest that the activity was memorable and helps develop mastery.** a) Most students reported a deeper understanding of all the devices built. b) All students reported no additional confusion arising from the activity, with a single notable comment that the student felt confused about more traditional fabrication methods. c) The diversity of comments shows that each student valued different aspects of the activity. This suggests the activity supports individualized development of mastery that helps fill key knowledge gap.

**Instructor observations and impressions**

In addition to the empirical evidence of learning, readers interested in adopting this activity may be interested in our learning related observations as we facilitated the activity. Overall, student displayed positive and enthusiastic reactions to the activity. Body language and focused discussion suggested that students engaged with the activity and predominantly completed the activity in an independent fashion, with an occasional clarifying question. Students also enjoyed experimenting with their completed systems, with one team creating a mini-competition to determine which individual could apply more pressure (and fluid) in their T-mixer, and thus have a channel dominated by a single colored fluid, rather than a split stream. Another team experimented with different pressures in the droplet generator and observed that a certain ratio of flows was needed to create droplets instead of two streams of oil bisected by colored water.

Most teams were able to construct functional devices with the parts initially distributed, only 2 devices of 20 were initially misassembled. When a spare-kit was provided, all teams were able to construct the devices. The most common challenges associated with assembling the devices were correctly orienting and aligning the layers, and ensuring that the layers were completely flat and able to provide a water-tight seal. Providing an opportunity to misassemble a device and quickly learn receive visual feedback was a key aspect to the learning activity overall. When provided with a spare kit and the opportunity to view a correctly assembled device prepared ahead of time, all teams were able to assemble all devices.

We observed that students adopted many different strategies for completing the device assembly and testing, including using hands instead of tweezers, working together, and/or taking turns experimenting/observing (Fig 6). All strategies proved to be equally successful in enabling students to complete the activity. Some teams initially misconnected the syringes with accompanying fluids (colored water, oil, or air) and associated ports, but were able to quickly change the connections after observing that the device did not operate as described in the handout. Overall, this was viewed as an unexpected asset as it provided students an opportunity to experiment, receive immediate feedback, and adjust without detrimentally affecting the device operation or results.

In future iterations of this activity, we plan on making several changes. First, while the time allotted for the activity was sufficient (2 sessions at 1.25 hours each), students verbally communicated that double this amount of time for additional exploration of the device function would have been welcome. Second, we will replace the coverslip bottom of the devices with more substantial, thick glass slides as several coverslips inadvertently broke during assembly. A best practice that we will continue is having a backup kit for each device for every 2-3 teams. We also recommend having one to two fully assembled devices of each type that can provided to teams for visualization and experimentation if necessary.

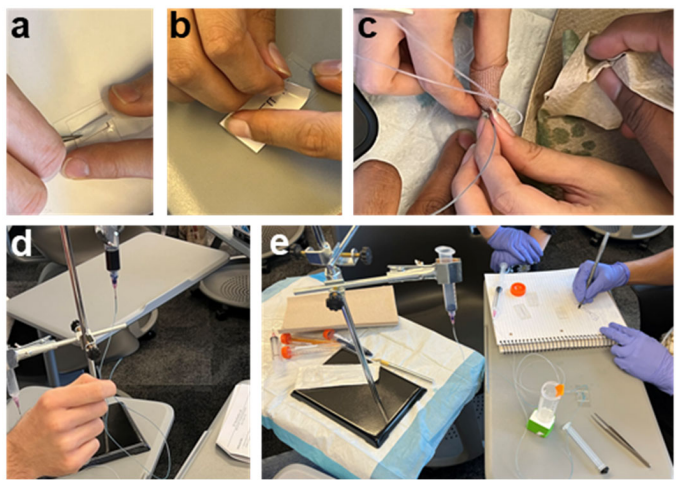

**Fig 6 Students adopt different successful strategies to build and test sticker microfluidics.** Different individuals preferred to build the microfluidics a) with and b) without tweezers. c) Some students preferred a team approach to attach fluidic ports whereas other teams d-e) preferred to take turns. Regardless of the approach taken, there was 100% success with regards to building and testing the devices. Key to this success rate was having several spare reserve device kits enabling students to have multiple attempts if needed at assembly.

**Study limitations**

There are several limitations with the current study. First, we have only been able to test student learning in a cohort of 11 students. While this is enough to confirm that the activity confers benefit, additional testing could include control groups and different interventions to further investigate the impact of microfluidic activities on student mastery. Second, our activity was administered in the broader context of a class that taught concepts related to microfluidic devices. As such, it is unclear if the activity alone teaches students important concepts or if student improvements are from including the activity in the context of a lecture-based course. Although the activity did include concepts that were not taught in the class, the class itself may have provided foundational structure and concepts relevant to student learning that improved student overall understanding.

**CONCLUSIONS**

Our work is unique in the context of the many excellent educationally focused papers describing microfluidic and microdevice activities. Prior to this work, the most complex structure constructed or tested in a classroom setting was a multi-layered serpentine channel [20]. This serpentine channel was important in that it likely helps students understand a multi-layered construction and visualize bubbles created in an upstream bubble generator, although the conferred benefit to students wasn't explicitly measured. In all prior microfluidic educational activities including the aforementioned serpentine channel, the "active" device layer is contained in a single microfluidic layer. Our work now enables students to build and test geometrically complex, multilayered devices, such as F-mixers and moving valves in an economical manner, with no external equipment, and a time commitment of two class periods. Unlike previous approaches, the function of the devices that students construct in our activity depends on the correct assembly of multiple layers, and more closely mimics the multilayered microdevices used in microfluidic research and commercial ventures.

Our data shows that our activity promotes significant gains in student knowledge and mastery for microfluidic physics and manufacturing. One unique aspect to our approach is that it helps students develop better spatial reasoning skills after completing this activity as evidenced by an improved ability to draw 2D top view and cross-sectional drawings of 3D microdevices that could achieve a desired outcome. This skill is key to both designing new microfluidics and communicating how to recreate another microfluidic as evidenced by the common practice of including cross-sectional drawings of microdevices in research reports. In addition, by providing quantitative data, our work also helps motivate the use of microfluidic learning activities in the classroom setting and addresses the overall lack of data in the field as a whole. Only two other studies have measured student learning [19], [21], and only one specifically examined student knowledge on microfluidics from completing an activity [21]. While this study did an excellent job of promoting student mastery, our approach is both faster and more economical.

.

Finally, one key advantage to our approach is that it utilized double-sided tape to create microfluidic devices, which is a well established tool in the microfluidics community that has been used to create a variety of economical devices as exemplified here [28]–[30]. Hence, the activity is both a pedagogical tool that facilitates student mastery and also a research tool that can be used for microfluidic and microdevice creation. Since our activity is highly portable, and could be performed in any classroom worldwide, our activity has the potential to democratize the education of miniaturization science. This could empower scientists in low-resource settings to leverage powerful new point-of-care technologies for unmet needs and address the gap in access to quality laboratory medicine [31]–[34]. "Flipped" training activities, such as that described in our report, provide novel opportunities for participatory research that challenges long-standing power relationships existent within the framework of Western knowledge generation [35]. Such research would allow local scientists who have firsthand knowledge of both the needs and challenges related to working in their environment to drive the design of new microfluidics and apply advancements in miniaturization science to critical healthcare problems affecting their patient population, which helps to resolve a broader longstanding educational challenge [36]. Importantly, the low-cost nature of this exercise allows for trial-and-error learning in microfluidic device construction, which is not feasible with high-cost fabrication techniques in resource constrained settings. Improved education in miniaturization science and microdevice fabrication could foster creation of unique, tailored solutions to common supply chain and stockout challenges faces in low-resource settings, particularly when laboratory testing is reliant on access to prefabricated, complex and proprietary consumables.

**ACKNOWLEDGEMENTS**

Financial support provided by R01HL155330 & K25HL141636 to DRM, F31HL160210-01 & American Society of Hematology Minority Medical Student Award Program Fellowship to OO, and NSF GRFP to CAL and AD. DRM thanks CRD for helpful discussions.

**AUTHOR CONTRIBUTIONS**

PD, SS, TH, & DRM. planned the work. PD, CAL, AD, & DRM performed the experiments and analyzed data. PD, JJW, SS, TH, & DRM interpreted the data and wrote the manuscript.

**DECLARATIONS**

No conflicts of interest. GT IRB reviewed and determined this study to be exempt. All students consented to participate**.** Data available upon request